\documentclass[
 aip, jcp,
 amsmath,amssymb,
preprint,%
 reprint,%
longbibliography
]{revtex4-1}

\usepackage{graphicx}
\usepackage{dcolumn}
\usepackage{bm}

\usepackage{float}
\usepackage[utf8]{inputenc}
\usepackage[T1]{fontenc}\usepackage{amsmath}

\usepackage[hidelinks]{hyperref}
\usepackage{comment}
\usepackage{amssymb}
\usepackage{xcolor}
\usepackage{float} 
\usepackage{dsfont}

\begin{document}

\title{Counting, Computing, and Pattern Recognition with Self-Assembling Non-Reciprocal DNA Tiles \vspace{10pt}
}

\author{Tim E. Veenstra}
\affiliation{Soft Condensed Matter \& Biophysics, Debye Institute for Nanomaterials Science, Utrecht University, Princetonplein 1, 3584 CC Utrecht, The Netherlands}
\author{Ren\'{e} van Roij}%
\affiliation{Institute for Theoretical Physics, Utrecht University,  Princetonplein 5, 3584 CC Utrecht, The Netherlands}
\author{Marjolein Dijkstra}
\affiliation{Soft Condensed Matter \& Biophysics, Debye Institute for Nanomaterials Science, Utrecht University, Princetonplein 1, 3584 CC Utrecht, The Netherlands}

\date{\today}
\pacs{}
\begin{abstract} Harnessing the intrinsic dynamics of physical systems for information processing opens new avenues for computation embodied in matter.  Using simulations of a model system, we show that assemblies of DNA tiles capable of self-organizing into multiple target structures can perform basic computational tasks analogous to those of \textit{finite-state automata} when equipped with  programmable non-reciprocal interactions that drive controlled dynamical transitions between these structures. By establishing design rules for multifarious self-assembly while budgeting the energy input required to drive these non-equilibrium transitions, we demonstrate that these systems can execute a wide variety of tasks including counting, computing modulo functions, and recognizing specific input patterns. This framework integrates memory, sensing, and actuation within a single physical platform, paving the way toward energy-efficient physical computation embedded in materials ranging from DNA and enzymes to proteins and colloids. 
\end{abstract}
\maketitle


Biological systems display remarkable non-equilibrium behavior,  continuously consuming  and dissipating energy to sustain life. This energy drives diverse processes, from cargo  transport  by kinesin proteins, to the flagellar motion powered  by  proton transport,  to ion pumps maintaining  membrane potentials. Many biological materials feature building blocks that selectively bind to form complex, functional structures. Striking examples are protein condensates, dynamic molecular entities  that  assemble and disassemble to perform specialized cellular functions. Advances in  bio-inspired nanotechnology have extended this principle to synthetic systems, using nucleic acids that interact via Watson-Crick-Franklin hybridization.\cite{Seeman2017}  Such interactions enable addressable  structures, where each building block occupies a unique position within the target structure by binding selectively to designated neighbors. This principle has enabled the design and assembly of a wide range of structures using DNA tiles,\cite{Seeman1982,Murugan2015,Osat2022,Evans2024} DNA origami,\cite{liu2016diamond,tian2020ordered,sigl2021programmable,liu2024inverse} and DNA-grafted colloids.\cite{Biancaniello2005, Nykypanchuk2008, mcmullen2022self, Rogers2015, Rogers2016, Chakraborty2017, He2020, Moerman2023} These addressable systems have enabled  applications in  drug delivery, biological sensing, catalysis, and photonics.\cite{Rogers2016}  

Building blocks need not belong to a single structure; identical components can participate in multiple assemblies, as seen in proteins forming  distinct  complexes.\cite{Sartori2020} This  multifariousness enables the self-assembly of several target structures from a shared library of building blocks.\cite{Murugan2015, Bupathy2022, Metson2025a} Each target structure corresponds to a minimum in a high-dimensional free-energy landscape, acting as attractors\cite{Keim2019} and making multifarious self-assembly analogous to associative memory in Hopfield networks.\cite{Hopfield1982} Here we will see that even richer behavior can emerge when non-equilibrium dynamics is introduced.

A particularly exciting class of non-equilibrium phenomena involves non-reciprocal interactions, where action-reaction symmetry is  broken. Such interactions violate time-reversal symmetry and convert energy irreversibly into useful work, leading  to rich dynamical behavior.\cite{Fruchart2021, Barrat2023, Dinelli2023} Non-reciprocity has been exploited to create predator-prey droplets,\cite{Meredith2020, YLiu2024} active solids,\cite{Brandenbourger2019,Veenstra2025} motile particle clusters,\cite{Navas2024} and diverse patterns.\cite{Saha2020, Rana2024, Huang2024} Crucially, it enables systems to escape kinetic traps and cross energy barriers,\cite{Navas2024, Osat2024, Osat2022} a property leveraged in multifarious assemblies. When building blocks exhibit non-reciprocal interactions, transitions between distinct target structures can occur in a directed and controlled manner.\cite{Osat2022} By breaking detailed balance, non-reciprocity allows  transitions between a sequence of energy minima, which effectively endows systems with ``shape-shifting'' behavior such that the next target structure nucleates and grows out of the previous one. 
\cite{Metson2025b} 

Here, we leverage non-reciprocal interactions not only to drive transitions between multifarious target structures, but also to design information-processing devices that accept inputs, process them, and generate outputs. 
Specifically, we develop finite-state automata for basic computations in a Brownian system with well-defined and controllable transitions between discrete states. Our approach employs multifarious self-assembly, in which a shared library of building blocks can self-organize into multiple target structures, which we use as  distinct states of a finite-state machine.\cite{Sipser2012, Kwakernaak2023, JLiu2024} Subsequently, we exploit non-reciprocal interactions to induce transitions between the different target states. We establish design rules to control these non-reciprocal transitions while budgeting the energy input (fuel) required to drive the non-equilibrium dynamics. Using this framework, we demonstrate systems capable of counting, computing modulo functions, and recognizing input patterns. 
\begin{figure*}[t!]
    \centering
    \includegraphics[width=\textwidth]{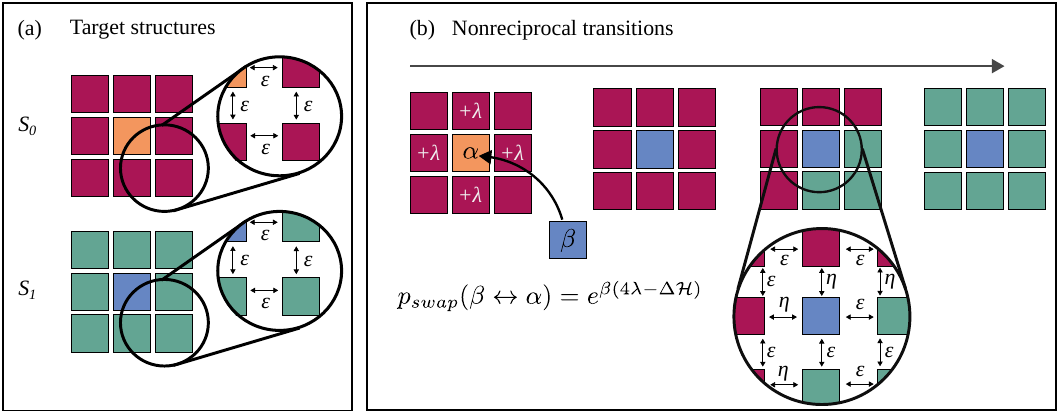}
    \caption{(a) Schematic representation of two target structures, labeled $S_0$ and $S_1$, both composed of a library that consists of nine different building block species. Particle colors in these sketches are arbitrary and only intended to distinguish different target structures, particles of the same color represent different particle species. Neighboring particles within the target structures bind with a strength $\varepsilon$ as shown in the insets. (b) Schematic of a few of the intermediate steps during the transition $S_0\rightarrow S_1$ as facilitated by the non-reciprocal interactions $\lambda$ and the reciprocal interactions $\eta$ between neighboring building blocks in the initial and the subsequent structure. Here we show that the particle of species $\beta$ from target structure $S_1$ replaces one of species $\alpha$ from $S_0$, driven by the availability of four times the non-reciprocal energy contribution $\lambda$ in the swap rate $p_{swap}$, as defined in Eq.(\ref{eq:swaprate}).  The neighboring particles from two consecutive target structures (here $S_0$ and $S_1$) are bound together during the transition with bond strength $\eta$ to stabilize the intermediate structures, as shown in the inset.  }
    \label{fig:interactions}
\end{figure*}

\section*{Distinct states and Controlled Transitions}
We consider a multicomponent mixture of $N_s$ DNA tile species on a three-dimensional cubic lattice, each with directional binding sites on its four faces in the $xy$-plane (Methods). These programmed directional interactions are designed to promote  self-assembly  into one of 
$m$ distinct two-dimensional multifarious target structures $S_0, S_1, \dots, S_{m-1}$ (Fig.\ref{fig:interactions}(a)). Each target structure comprises of   
$N_t=18\times 16=288$ tiles drawn from a shared library of species. The stability of these structures arises from  directional bonds of energy  
$-\varepsilon=-6.8 k_BT$ between neighboring tiles. 

To enable controlled transitions between target structures,  we introduce non-reciprocal “swap moves” between DNA tile species $\alpha$ and $\beta$ (Fig.\ref{fig:interactions}(b)).\cite{Osat2022} When $\alpha$ is stably bound in the low-energy state of structure $S_\ell$,  swaps are rare. To overcome this, an energy input $\lambda>0$ —representing external fuel, enzymatic activity, or concentration gradients— is used to break an $\alpha$–$\gamma$ bond in $S_\ell$ and form a new  $\beta$–$\gamma$ bond. This occurs only when $\beta\in S_{\ell+1}$ is adjacent to $\gamma\in S_\ell$ (Methods).

We found that stabilizing intermediate structures containing  particles from both $S_{\ell}$ and $S_{\ell+1}$ is essential. Without this stabilization, $S_{\ell +1}$ particles or clusters detach from  $S_\ell$ before the transition to $S_{\ell+1}$  completes. To prevent this, we introduce an  inter-target binding energy $-\eta<0$ between neighbouring DNA tiles  $\alpha$ and $\beta$ that are adjacent in both preceding and subsequent  structures, ensuring reciprocity (Fig.\ref{fig:interactions}(b)). 

\begin{figure}
    \centering
    \includegraphics[width=\linewidth]{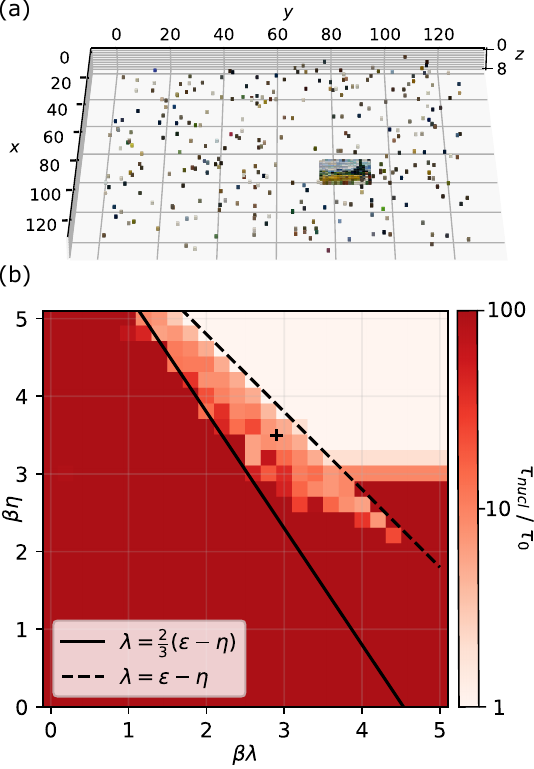}
    \caption{(a) Typical configuration of $576$ different DNA tiles, modeled as unit cubes with a fixed orientation and endowed with specific directional nearest neighbour binding of the four faces in the $xy$-plane. The simulation box is a $128\times128\times8$ cubic lattice. A complete  two-dimensional target structure of 18$\times$16=288 DNA tiles has successfully assembled, and one other particle library of 18$\times$16 different DNA tiles is dispersed in the simulation box. Each species of DNA tiles is assigned a unique color, chosen such that the self-assembled target structure resembles Vincent van Gogh's painting \textit{Wheatfield with cypresses}. (b) Heatmap of the nucleation time $\tau_{nucl}/\tau_0$ as a function of the inter-target  interaction strength $\beta \eta$ and non-reciprocity value $\beta\lambda$, along with the lines denoting the parameter regime where the $S_0 \rightarrow S_1$ transitions occur reliably.}
    \label{fig:phase-diagram}
\end{figure}

Although our DNA tile model is simple, it involves many independent parameters. These include the number of species $N_s$, the number of tiles per target structure $N_t$, and the number of target structures $m$, as well as the binding energies $\epsilon$ and $\eta$, the fuel energy $\lambda$, and the thermal energy $k_BT=1/\beta$. We use simulations to identify parameter sets suitable for constructing finite-state automata from self-assembling DNA tiles.
We define the tile-tile interactions to support two target structures $S_0$ and $S_1$ with non-overlapping libraries. Each species is assigned a color, such that the fully self-assembled structures $S_0$ and $S_1$ resemble Van Gogh's  {\em Wheatfield with Cypresses}  and Vermeer's {\em The Milkmaid} painting, respectively. 
To study the transition $S_0\rightarrow S_1$, we perform Monte Carlo (MC) simulations for a range of stabilizing interaction strengths $\beta \eta \in [0,5]$ and non-reciprocity values $\beta \lambda \in [0,5]$. Each simulation starts  from a state  resembling  the snapshot in Fig.\ref{fig:phase-diagram}(a), with a fully assembled  $S_0$ and the DNA tiles  for  $S_1$ freely dispersed in the  fluid. We determine the nucleation time $\tau_{nucl}$, i.e. the time that  the transition  to $S_1$ has occurred (Methods).
Fig.\ref{fig:phase-diagram}(b) shows a heat map of $\tau_{nucl}/\tau_0$ as a function of $\beta\eta$ and $\beta\lambda$ with $\tau_0=5\cdot 10^5$ MC sweeps.
For low $\beta\eta$ and $\beta\lambda$ (dark red), no transitions occur, confirming that both active driving ($\lambda>0$) and stabilization ($\eta>0$) are needed.
For high values (white), transitions are nearly instantaneous; when $\lambda\gg\varepsilon-\eta$, the nucleation barrier vanishes, causing chaotic behavior.\cite{Osat2022}
Controlled transitions with finite nucleation times ($\sim10\tau_0$) occur in the intermediate regime $2(\varepsilon-\eta)/3 \le \lambda \le \varepsilon-\eta$.
Transitions fail for $\beta\eta\lesssim2.5$ or when $\lambda\ge\varepsilon-\eta$ and $\beta\eta<3$, as weak stabilization leads to melting.
We therefore choose $\beta\epsilon=6.8$, $\beta\lambda=2.9$, and $\beta\eta=3.5$ (black cross) as optimal parameters for reliable $S_0\rightarrow S_1$ transitions.

\begin{figure*}[]
    \centering
    \includegraphics[]{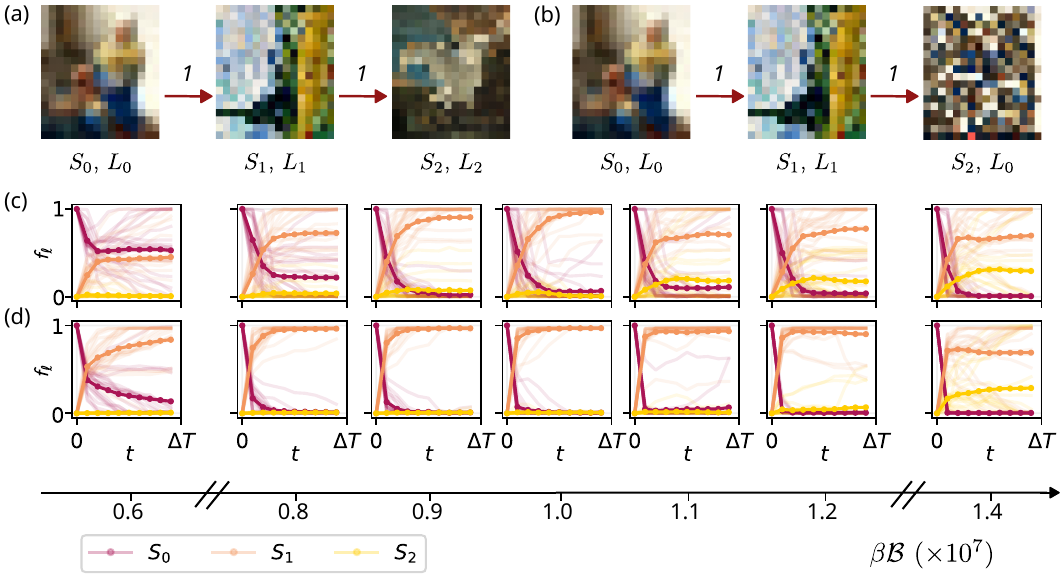}
    \caption{Transition paths with three target structures using (a) three unique particle libraries or (b) two particle libraries, where  structures $S_0$ and $S_2$ share most of their particle species. Each particle species has a unique color, chosen such that $S_0$, $S_1$ and $S_2$ resemble the paintings \textit{The milkmaid} by Johannes Vermeer, \textit{Wheatfield with cypresses} by Vincent van Gogh, and \textit{The threatened swan} by Jan Asselijn, respectively. In panel (b)  $S_2$ uses the same particle library as $S_0$, resulting in a scrambled version of  \textit{The milkmaid}. In both systems, the  transitions $S_0\rightarrow S_1$ and $S_1\rightarrow S_2$ are triggered by the same input label and are therefore simultaneously accessible. (c,d) The fraction of bonded neighbors $f_\ell$ for all structures $S_\ell$ during a single time window plotted as a function of time (solid lines), for various values of non-reciprocity budget $\mathcal{B}$, averaged over 21 individual simulations (transparent lines).   Panel (c) corresponds to the system shown in (a), while panel (d) shows the results for the multifarious system illustrated in (b).}
    \label{fig:budget-sequence}
\end{figure*}

\section*{Controlled Input-Triggered Transitions}
The next step toward a Brownian finite-state machine is the extension to DNA tile libraries that assemble into multiple target structures. Key challenges to address include designing transitions that respond to  input signals, ensuring transitions to full completion, and avoiding premature ones. We address these using discrete MC time intervals during which  input pulses are given in terms of a set of time-dependent non-reciprocities $\lambda_{\ell k}(t)$, and alternating libraries of multifarious DNA tile species (see Methods). Here, we focus on two types of input pulses. We perform MC simulations starting from a fully assembled  structure $S_0$ at $t=0$.  Transitions are analyzed over discrete time windows $t\in[t_n,t_n+\Delta T]$, where  $t_n=n\Delta T$ for integer $n\geq 0$. The time window $\Delta T=100\tau_0$ is sufficiently long  to complete the transition $S_\ell \to S_{\ell+1}$. A transition $S_\ell\rightarrow S_k$ is activated by a non-reciprocal energy $\lambda_{\ell k}(t_n)\neq 0$. To avoid unintended transitions, $\lambda_{\ell k}(t)$ decreases gradually after $t_n$ due to a finite fuel budget $\mathcal{B}$. In systems with multiple target structures,  overlapping particle libraries are used for alternating structures, ensuring depletion of DNA  tiles needed for later structures and thus suppressing premature nucleation. 

Fig.~\ref{fig:budget-sequence} illustrates the effectiveness of combining a finite non-reciprocity budget with alternating libraries. We examine two systems with three target structures ($S_0$, $S_1$, $S_2$) and two input-triggered transitions ($S_0\rightarrow S_1$ and $S_1 \rightarrow S_2$).
As before $S_0$ and $S_1$ are representations of Vermeer and Van Gogh paintings, assembled from non-overlapping libraries $L_0$ and $L_1$ (288 species each).
The difference lies in $S_2$: in Fig.~\ref{fig:budget-sequence}(a), it is  Asselijn’s {\em The Threatened Swan} built from a distinct library $L_2$; in Fig.~\ref{fig:budget-sequence}(b), $S_2$ largely reuses $L_0$,  sharing 286 of 288 species with $S_0$,  forming a scrambled Vermeer. For both systems, we run 21 simulations over  $t\in[0,\Delta T]$, starting from  $S_0$ with default non-reciprocity $\lambda_{01}(0)=\lambda_{12}(0)=\lambda_0$ with $\beta\lambda_0=2.9$ (representing input “1”) and a range of budgets $\mathcal{B}\in [0.6,1.4]\cdot10^7$ in units of $k_BT$. The target transition is $S_0\rightarrow S_1$, so any final structure other than $S_1$ at $t=\Delta T$ is undesired. Figs.~\ref{fig:budget-sequence}(c,d) show the $t$-dependence of the fraction of bonded neighbors for $S_0$ (red), $S_1$ (orange), and $S_2$ (yellow). For the non-multifarious system (c), low budgets ($\mathcal{B}=0.6\cdot10^7$) often yield incomplete transitions, while higher budgets reduce $S_0$ remnants but trigger premature $S_2$ nucleation. By contrast, in the multifarious system (d) the complete transition $S_0\rightarrow S_1$ is  highly reliable across a wide range of $\mathcal{B}$ while $S_1\rightarrow S_2$ is suppressed.  As Fig.~\ref{fig:budget-sequence}(d) shows, $S_1$  dominates  at $t=\Delta T$ in at least 85\% of simulations for $\mathcal{B} \in [0.6, 1.2]\cdot10^7$, reaching 100\% success  for $\mathcal{B} \in [0.8, 1.0]\cdot10^7$.

\section*{Finite-State Automata using DNA Tiles}

\begin{figure}[t!]
    \centering
    \includegraphics[width=\linewidth]{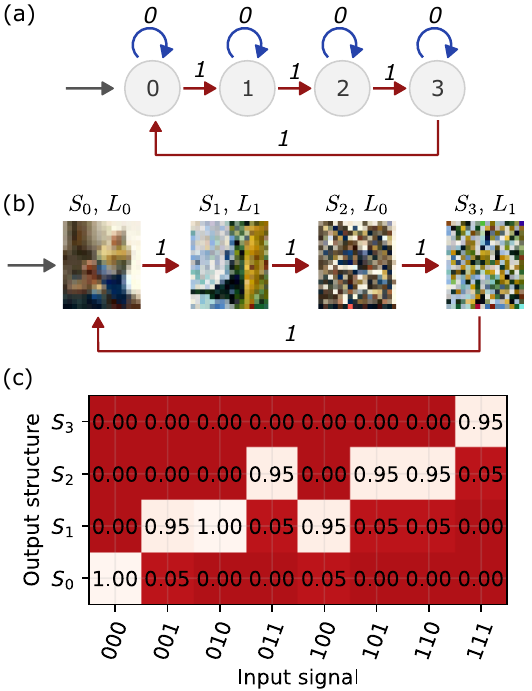}
    \caption{(a) Transition graph of a finite-state automaton for counting the number of ``1''-bits in binary inputs 000 through 111, in (a) a conventional abstract representation and in (b) our proposed physical realization using four multifarious structures $S_\ell$ for $\ell\in\{0,1,2,3\}$ with alternating particle libraries $L_i$ for $i\in\{0,1\}$ and directed transitions $S_{\ell}\rightarrow S_{\ell+1}$ triggered by non-reciprocal interactions. After an input of $\ell$ chemical fuel pulses (see text for details), the final structure is (ideally) $S_\ell$ such that this is a physical device that can count up to three. (b) The fidelity of the Brownian automaton of Fig.~\ref{fig:sketch-counting}(b), showing the fraction of assembled structures $S_\ell$  of all seven inputs 001 through 111 (realized physically as fuel-budget pulses) as obtained from 21 simulations for each input. 
    For all inputs, the correct result, corresponding to the number of ``1'' bits in the input, is obtained in at least 95\% of the simulations. 
    }
    \label{fig:sketch-counting}
\end{figure}

    


Having established distinct states and well-controlled transitions using a limited fuel budget and alternating particle libraries, we can now construct finite-state machines for sequential information processing.

Counting (to three)---Counting is a fundamental  computational operation and a natural starting point for our multifarious self-assembly model of Brownian DNA tiles. We design a system that receives a binary input sequence  and counts the number of  ``1'' bits in the signal. For this proof-of-concept,  we limit the count  to values from zero to three (modulo 4), and consider all possible three-bit binary inputs from 000 to 111. Fig.~\ref{fig:sketch-counting}(a) shows an abstract finite-state automaton with four states (0-3). A transition to the next state occurs when the binary input is  ``1'', while input ``0'' leaves the state unchanged. Counting modulo 4 thus requires four distinct states, equivalent to  two bits of memory. Each state label represents the total number of ``1''-bits received, assuming  the system starts in state 0.   Fig.~\ref{fig:sketch-counting}(b) illustrates our physical implementation using  $N_s=582$ distinct DNA tile species that can assemble into four target structures $S_0$, $S_1$, $S_2$, and $S_3$, representing  output states $0-3$.   Unlike the earlier two-state system (with only $S_0$ and $S_1$), this design  is strongly multifarious: $S_0$ and $S_2$ share  library $L_0$ ($290$ species of which $286$ common), and  $S_1$ and $S_3$ share library $L_1$ ($292$ species of which $284$ common). $S_0$ and $S_1$ correspond to the Vermeer and Van Gogh paintings used previously, while $S_2$  and $S_3$ are scrambled versions of them, using the same  libraries. A snapshot of  state $S_0$ thus   closely resembles the 576-tile configuration in Fig.~\ref{fig:phase-diagram}(a). 

As shown above, controlled directional transitions between target structures can be induced by combining alternating particle libraries with a budget-limited fuel supply (Eq.(\ref{B})). We extend this approach by representing the $n$-th input bit ``1'' and ``0'' by setting $\lambda_{01}=\lambda_{12}=\lambda_{23}=\lambda_{30}=\lambda_0$ and $0$, respectively, at time $t_n$ with a budget $\mathcal{B} =1.0\cdot 10^7$. 
Using this algorithm, we perform 21 simulations for each of the eight three-digit binary inputs (000-111), starting from a fully assembled  $S_0$ with species from library $L_1$ dispersed in the fluid.  
As shown  in Fig. \ref{fig:sketch-counting}(c), the number of ``1'' bits   is correctly counted in at least 95\% of the simulations. Errors, occurring  in fewer than 5\% of the cases, correspond to  a single missed transition;  no  multiple missed  or  additional transitions are observed.  These results demonstrate that our  system functions as a finite-state automaton capable of counting  with high, though not perfect, fidelity ---thus establishing a  basis for the design of more  complex  computational automata.

\begin{figure*}[t!]
    \centering
    \includegraphics[width=\linewidth]{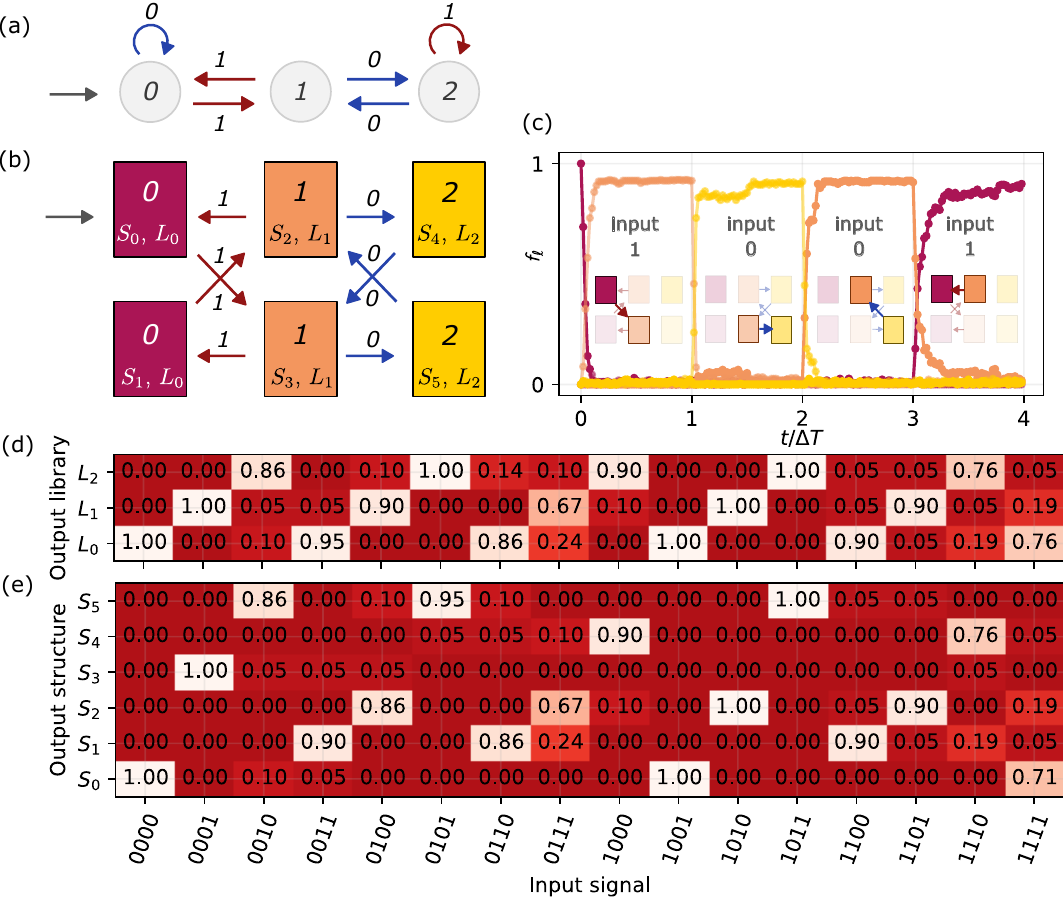}
    \caption{Transition graph of a finite-state automaton for calculating the modulo three of a binary input, in (a) a conventional representation and in (b) our equivalent implementation using multifarious structures with non-reciprocal transitions. Here states labeled $L_i$ are congruent with $i \bmod 3$. The target structures are represented by solid  rectangles, and colored according to their particle library because the scrambled paintings of libraries $L_0$ and $L_2$ (as shown in Fig. \ref{fig:budget-sequence}(a)) are difficult to distinguish by eye. (c) The fraction of bonds of each target structure $f_\ell$ as a function of time. Typical trajectory of the composition at binary input sequence $1001$ (decimal 9). The input and the transitions graph representation of the specific transition that occurs in each time window is shown overlaid with the figure. (d-e) The fraction of simulations which resulted in (d) the 3 different libraries and (e) the 6 possible output structures of the finite-state automaton that calculates modulo 3 of a binary input signal, for all 16 binary input signals between 0000 and 1111. }
    \label{fig:binmod3}
\end{figure*}

Modulo computation---We now design  a Brownian finite-state automaton that computes $i\bmod3$ for a binary input  $i$. This task is    more challenging than  counting  ``1''s, as it requires the automaton to track the precise bit order.  The complexity of this  computation is evident in the schematic   finite-state automaton (Fig.~\ref{fig:binmod3}(a)).  Appending a ``0'' to a binary number doubles its value, affecting the  modulo accordingly. State ``0'' remains  unchanged ($(2\times 0)\text{mod}~3 =0$), while states 1 and 2 switch ($(2\times1)\text{mod}~3=2$ and $(2\times2)\text{mod}~3=1$)). Appending a ``1'' doubles the number  and adds 1, causing transitions $0\leftrightarrow1$, but not between 1 and 2. 

A key challenge is that the same input can drive   opposite transitions: input ``1'' triggers both   $0 \to 1$ and  $1 \to 0$, while  input ``0'' triggers both $1 \to 2$ and $2 \to 1$. Such transitions resemble detailed balance---fundamentally at odds with our design principle of directed, non-reciprocal transitions  driven by irreversible fuel consumption. We tackle this problem by leveraging the multifarious nature of our design to effectively ``double'' the number of target structures while  reusing the same particle libraries. As shown in Fig.~\ref{fig:binmod3}(b), the  automaton comprises  six target structures ($S_0-S_5$) and three libraries ($L_0,L_1,L_2$):  $S_0$ and $S_1$ share  $L_0$,  $S_2$ and $S_3$ share $L_1$, and $S_4$ and $S_5$ share $L_3$. When initialized from $S_0$, this automaton is equivalent to that in Fig.~\ref{fig:binmod3}(a) but avoids  opposite transitions triggered by the same input, making it compatible with  our non-reciprocal, fuel-driven transition strategy.  This design also keeps the total number of species nearly constant through shared libraries, though with slightly more intricate species architectures. The result of the computation,  $i\bmod3$, is  encoded in the library label $n=0,1,2$ of $L_n$ of the final structure. The particle libraries $L_0$, $L_1$, and $L_2$ contain $290$, $292$, and $295$ species, respectively. As  before, simulations start from a fully assembled $S_0$. We perform 21 simulations for each of the 15 possible 4-bit input sequences (0000-1111). Binary inputs (``0'' and ``1'') are encoded  as external pulses  that trigger the corresponding  transitions ($\lambda_{24}=\lambda_{35}=\lambda_{52}=\lambda_{43}=\lambda_0$ for ``0'', and $\lambda_{03}=\lambda_{31}=\lambda_{12}=\lambda_{20}=\lambda_0$ for ``1'') at the start of each time window  $\Delta T$, with a  fuel budget $\mathcal{B}=1.2 \cdot 10^{7}$. We record the time-dependent fraction of satisfied bonds $N_\ell$ for each target structure $\ell$. A fully formed $S_\ell$ yields $N_\ell \approx 1$, while all others $N_{\ell'}$ ($\ell'\neq \ell$)  remain near zero. A representative trajectory for   input sequence 1001 (decimal 9) is shown in Fig. \ref{fig:binmod3}(c). Since $9\bmod3=0$, the final state should correspond to library $L_0$. During the first time window ($0<t<\Delta T$; input ``1''), the system transitions from  $S_0$ to $S_3$, reaching  $N_3>0.9$. Over the next two windows ($\Delta T<t<3\Delta T$; inputs  ``00''), it progresses via $S_5$ to $S_2$, maintaining high bond fractions. In the final window ($3\Delta T < t < 4\Delta T$; input ``1''), it correctly returns to $S_0$ associated with  $L_0$. 

Fig. \ref{fig:binmod3} summarizes the final-state distributions of  $S_\ell$ and  libraries $L_i$, averaged  over 21  simulations per input sequence. The correct structure---defined as  having at least half  the bonds of the intended target---appears in most simulations, with an average success rate of about   89\%. The statistics for the final library yields  similar or slightly higher accuracy, as imperfect structures can still belong to the  correct library. Overall, the Brownian finite-state automaton  computes $i\bmod 3$ correctly for most  inputs  from 1-15, though accuracy declines  for  longer inputs ($i\gg15$) due to cumulative transition errors. 


\begin{figure*}[t!]
    \centering
    \includegraphics[width=\linewidth]{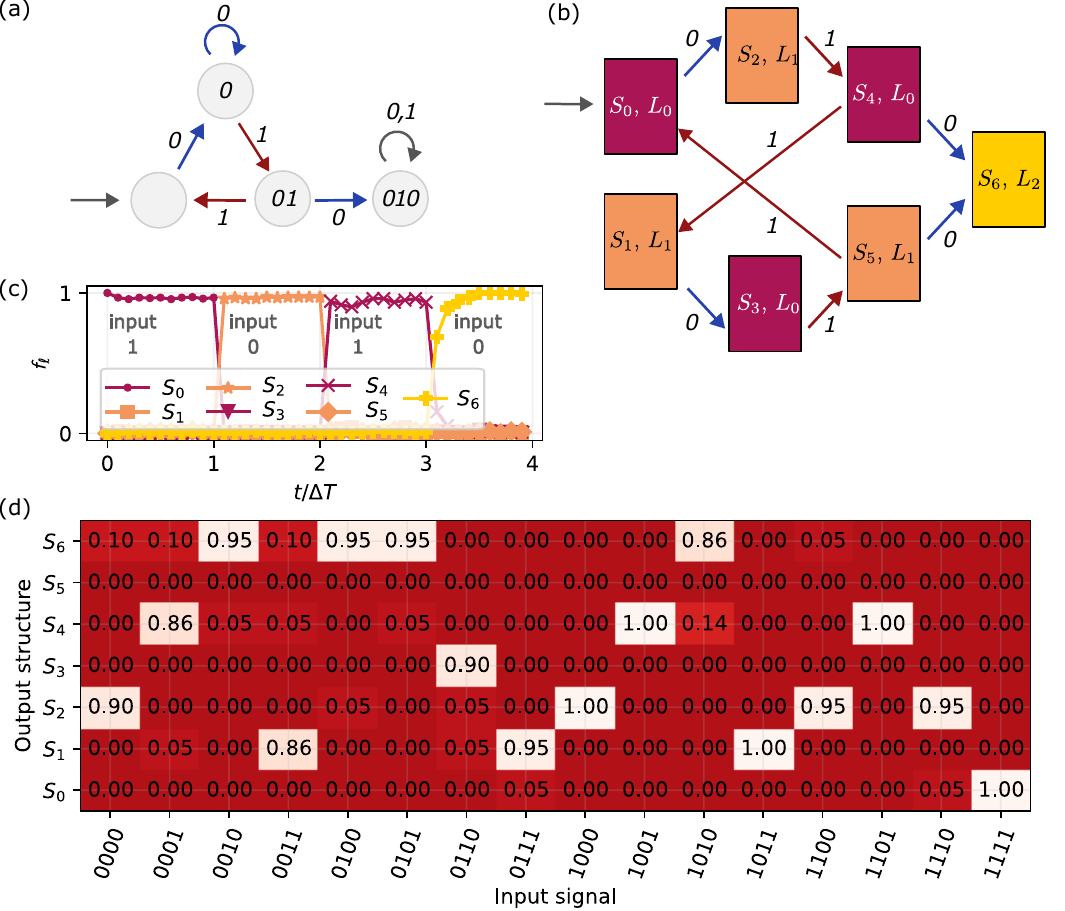}\\

    \caption{(a) Schematic finite-state automaton that recognizes the pattern *010* (or equivalently *101*, by swapping the input labels) at any point in an arbitrarily long input sequence, in (a) a conventional representation and in (b) our equivalent implementation using multifarious target structures and non-reciprocal transitions. The target structures are colored according to their particle library. (c) The fraction of bonds of each target structure $f_\ell$ as a function of time for the typical trajectory of the binary input sequence $1010$. The input at each time window is shown overlaid with the figure. (d) The fraction of simulations which resulted in the 7 possible output structures of the finite-state automaton that recognizes an input sequence containing *010*, as a function of the binary input signal.}
    \label{fig:acceptor}
\end{figure*}

Input pattern recognition---Finally, we design a Brownian finite-state automaton that recognizes a specific pattern within an input sequence. Such functionality could enable sensing applications, such as  detecting the presence or absence of a given sub-pattern. 
The transition graphs used previously are unsuitable for this task, as they lack  an acceptor (or terminal) state---one that can be entered but not exited. Such a state is essential to  signal  successful pattern recognition, regardless of any subsequent input. 
Fig. \ref{fig:acceptor} illustrates an automaton that detects the pattern  $*010*$  within an arbitrarily long  input. Panel (a) presents the  conventional finite-state diagram, while panel (b) depicts its physical realization using Brownian  DNA tiles. This implementation involves seven target structures ($S_0$-$S_6$), constructed from three libraries: $L_0$ (303 species;  assembling $S_0$, $S_3$, $S_4$), $L_1$ (305 species; assembling  $S_1$, $S_2$, $S_5$), and $L_3$ (288 species; assembling $S_6$). In total, the simulation involves  $N_s=896$ distinct DNA tile species. The design embeds a dead-end structure, $S_6$, that can be reached but not exited. The system begins from a fully assembled $S_0$ and evolves in response to binary input pulses (``0'' and ``1'') applied at the start of each  time window $\Delta T$, with  a fuel budget ${\cal B}=1.1\cdot10^7$. The automaton   reaches $S_6$  if and only if the input sequence contains the target pattern. Thus, whenever the pattern $*010*$ appears in the binary input sequence, the system transitions to the terminal structure $S_6$,  where it remains permanently. 

A representative trajectory for the input sequence $1010$, which  contains the target pattern $*010*$, is shown in Fig.~\ref{fig:acceptor}(c). During the first time window, no transition occurs since $S_0$ has no outgoing transition for input $1$. When the subsequent  sequence $*010*$ is applied,  the system correctly transitions to $S_6$, signaling successful recognition of the  pattern $*010*$. We perform 21 simulations for all 4-bit binary inputs (0000-1111), and  report the resulting distributions of final structures in Fig. \ref{fig:acceptor}(d). The terminal state $S_6$ appears with  high probability (86-95\%) for  the four inputs containing the pattern $*010*$, and with  low probability (0-10\%, including eight cases of exactly $0\%$) for  inputs lacking it. This clear separation between the two outcomes demonstrates the Brownian automaton can reliably perform pattern recognition. Of course, this approach is not limited to detecting the specific sequence $*010*$; finite-state automata can be designed to recognize  any desired bit pattern.\cite{Sipser2012} 

\section*{Conclusions}
This work demonstrates a physical computing system where  multifarious self-assembly,  combined with non-reciprocal transitions, performs computations as a finite-state automaton. A key challenge was preventing premature transitions, which can destabilize regular switching trajectories and induce chaotic behavior.\cite{Osat2022} Such  premature switching also undermines sequential information processing by reducing  control over the number of transitions per input. To overcome this, we introduced a non-reciprocity budget and imposed a design rule on the multifarious structures to ensure that each input triggers only a single transition---an essential requirement for finite-state automata. Under these constraints, the system can perform  tasks such as counting, computing moduli of binary numbers, and recognizing specific input patterns.

More generally, any system with a finite number of states and controlled transition mechanisms  can, in principle, perform these tasks. Finite-state automata are powerful models of  computation, and their realization in physical systems opens avenues  for designing adaptive, smart materials. This framework therefore opens new avenues for energy-efficient  physical computation, information processing, and sensing.


\bibliography{library}

\section*{Methods}
\subsection*{Potts-like lattice model for DNA tiles}

We consider a computationally efficient model for a multi-component system composed of many species of DNA tiles, where each DNA tile is represented as a unit cube with a fixed orientation that resides on a cubic lattice of unit spacing in three dimensions. Each DNA tile has a single binding site on each of the four faces in the $xy$-plane, which allows for directional bonding with selected nearest neighbors, 
while the top and bottom faces (oriented in the $z$-direction) are inert. 
The directional interactions in the $xy$-plane are designed (as detailed below) in such a way that the system can self-assemble into one of  $m$ distinct, multifarious target structures $S_0, S_1, \dots, S_{m-1}$. Each of these structures consists of $N_t$ DNA tiles and corresponds to a potential-energy minimum of the system. An illustrative example of eighteen species ($N_s=18$) and only two target structures ($m=2$), each composed of nine DNA tiles ($N_t=9)$, is shown in Fig. \ref{fig:interactions}(a). In principle, also multiple target structures can be composed from (largely) the same set of DNA tiles, which we refer to as a \textit{library} of DNA tiles.

This lattice model can conveniently be formulated as a Potts model, where each lattice site $i$ is either empty or occupied by at most one DNA tile. The state of lattice site $i$ can thus be represented as a Potts-like spin variable $s_i$, which is an $N_s$-dimensional vector with components $s_{i,\mu}$ where $\mu\in\{1,2,\cdots,N_s\}$ is the species index. We write $s_{i,\mu}=\delta_{\mu\alpha}$ in terms of the Kronecker-$\delta$ if site $i$ is occupied by a DNA-tile of species $\alpha\in\{1,2,\cdots,N_s\}$, and $s_{i,\mu}=0$ if site $i$ is not occupied by any DNA tile.

In our model, we consider only nearest-neighbor interactions, which allows us to write the interaction energy between two nearest-neighbor sites with states $s_i$ and $s_j$ as the vector-matrix-vector contraction $s_i\cdot J(d)\cdot s_j$.  Here, $J(d)$ denotes a  real-valued $N_s\times N_s$ nearest-neighbor interaction matrix, with components  $J_{\alpha\beta}(d)$. The discrete parameter $d\in\{\rightarrow, \uparrow, \leftarrow,\downarrow\}$ explicitly encodes the directional dependence of the interaction between species $\alpha$ and $\beta$,  distinguishing the four possible relative orientations of the two DNA tiles in the $xy$-plane: it matters for the directional bonding at fixed orientation whether a tile of species $\beta$ is to the right ($\rightarrow$) or left ($\leftarrow$) or behind ($\uparrow$) or in front of ($\downarrow$) a tile of species $\alpha$. Below we will construct the decomposition $J(d)=U(d)+{\cal V}(d)+\Psi(d)$. We start by discussing the directional intra-target bonding matrix $U(d)$, which specifies the interaction strengths  between specific species of DNA tiles. This matrix  is designed such that a multi-component mixture of these tiles tends  to self-assemble into one of $m$ distinct, multifarious, low-energy target structures 
$S_0, S_1, \dots, S_{m-1}$. Because the directional bonding potential must respect translational invariance, it satifies $U_{\alpha\beta}(d)=U_{\beta\alpha}(-d)$, where we understand that $(-\uparrow)=\,\downarrow$ and $(-\leftarrow)=\,\rightarrow$.  Next, we introduce the non-reciprocal interaction matrix ${\cal V}(d)$, which  not only initiates but also actively drives the system across energy barriers, enabling transitions  from a target structure $S_\ell$ to the subsequent structure $S_{\ell+1}$ (but not from $S_{\ell+1}$ back to $S_{\ell}$). Because this process is directional and irreversible, the interactions are non-reciprocal, and hence ${\cal V}_{\alpha\beta}(d)\neq {\cal V}_{\beta\alpha}(-d)$). Finally, we  introduce a weak inter-target binding matrix $\Psi(d)$, which satisfies reciprocity $\Psi_{\alpha\beta}(d)=\Psi_{\beta\alpha}(-d)$, and plays a crucial role in  preventing the dissolution of intermediate structures that form during transitions  between two consecutive target structures.

\subsection*{Non-reciprocal Interactions for Controlled Transitions}
\label{sec:interactions}
The DNA tiles are assumed to bind with energy $-\varepsilon<0$ if their facing sides are neighbors in at least one of the $m$ target structures. 

Two nearest-neighbor sites $i$ and $j$, with states $s_i$ and $s_j$, respectively, thus give a target-binding energy contribution $s_i\cdot U(d_{ij})\cdot s_j$ to the interaction energy, with the $\alpha\beta$ element of the intra-target binding energy matrix at relative orientation $d$ given by
\begin{equation}\label{U}
    U_{\alpha\beta}(d) = \begin{cases} 
                                -\varepsilon, & \text{if species $\alpha$ and $\beta$ bind with }\\& \text{relative orientation $d$ in a}\\& \text{target structure $S_\ell$;} \\
                                
                                0, & \text{otherwise}.
                            \end{cases}
\end{equation}

We note that, in addition to  confining the DNA tiles to a cubic lattice and restricting their orientations, we make the  simplification that all pairs of species $\alpha\beta$ that bind in a particular orientation $d$ have one and the same binding energy $-\varepsilon$. 
We also remark that the interaction matrix of Eq.(\ref{U}) is reciprocal, satisfying $U_{\alpha\beta}(d)=U_{\beta\alpha}(-d)$, which implies that $s_i\cdot U(d_{ij})\cdot s_j = s_j\cdot U(d_{ji})\cdot s_i$ is Hermitian, since $d_{ji}=-d_{ij}$.

Inspired by Osat \textit{et al.} \cite{Osat2022}, we introduce non-reciprocal swap moves between a DNA tile of species $\alpha$ at pre-swap lattice site $i$ and a DNA tile of species $\beta$ at pre-swap lattice site $j$. If such a swap is attempted when the DNA tile $i$ of species $\alpha$ is initially ``comfortably'' sitting  in a low-energy state within the target structure $S_\ell$, the swap generally increases the potential energy. As a consequence, this process is unlikely to be accepted in a Monte Carlo (MC) simulation, or to occur spontaneously by Brownian fluctuations under experimental conditions in an aqueous environment. To trigger and promote the transition from $S_\ell$ to $S_{\ell+1}$ across a potential barrier, 
such swaps can be actively facilitated by the energy input $\lambda>0$ of an external source, for example from  fuel molecules, concentration gradients, or enzymatic activity. This amount of energy can be irreversibly consumed to facilitate the breaking of an $\alpha-\gamma$ bond, where  $\gamma\in S_\ell$ is a neighbor of $\alpha$ in direction $d$, allowing it to be replaced by a new (potentially weaker or neutral) $\beta-\gamma$ bond. However, the consumption of the energy $\lambda$ is permitted {\em only} when the specific species $\beta\in S_{\ell+1}$ is involved that occupies a site neighboring to $\gamma\in S_{\ell}$, where ``neighbor'' is a well-defined concept for the equal-sized rectangular target structures of this study (but may need further specification for more general geometries of the target structures). We therefore define the non-reciprocal interaction matrix $\mathcal{V}(d)$ between two nearest neighbor tiles by
\begin{equation}\label{V}
    \mathcal{V}_{\beta \gamma}(d)=
    \begin{cases}  
        \lambda, & \text{if there exists an $\ell$ for which } \\ & \text{$\beta\in S_{\ell+1}$ and $\gamma\in S_\ell$ occupy } \\ & \text{neighboring sites in structures} \\ & \text{$S_{\ell+1}$ and $S_{\ell}$, respectively, in} \\ & \text{direction $d$;} \\ 
        0, & \text{otherwise.}    
    \end{cases}
\end{equation}
Here, $\lambda>0$ denotes the available fuel energy for the swap of a tile of species $\alpha\in S_\ell$ with $\beta\in S_{\ell+1}$, in the presence of a fixed neighbor particle of species $\gamma$ in  direction $d$.

One can explicitly verify that ${\cal V}_{\beta\gamma}(d)\neq {\cal V}_{\gamma\beta}(-d)$, reflecting the fact that the energy $\lambda$ does not stem from a particle-particle potential, but is instead  irreversibly consumed during the swap process $\alpha\leftrightarrow\beta$. This asymmetry is also reflected by the explicit dependence of Eq.(\ref{V}) on $\beta$ and its independence of $\alpha$. The dependence on $\alpha$ is implicit through the identity of its neighbors $\gamma$ in structure $S_\ell$, which would be different for the reverse swap $\beta\leftrightarrow\alpha$, since in that case $\gamma$ would represent the neighbors of $\beta$ in $S_{\ell+1}$ rather than those of $\alpha$.  The non-reciprocity of ${\cal V}$ manifests in the system's dynamics, facilitating the transition $S_{\ell}\rightarrow S_{\ell+1}$, while suppressing the reverse transition.

While DNA tiles of species $\beta\in S_{\ell+1}$ can thus gradually insert themselves into structure $S_{\ell}$ when the fuel energy $\lambda$ is available in suitable amounts (as we will see), it is crucial that intermediate structures, consisting of particles from both $S_{\ell}$ and $S_{\ell+1}$, an example of which is shown in Fig. \ref{fig:interactions}(b), are stabilized by energetic bonds. Without such stabilization, particles or clusters from $S_{\ell +1}$ would detach from those of $S_\ell$ before the transition to $S_{\ell+1}$ could be completed. We therefore introduce an additional inter-target interaction matrix $\Psi(d)$, which assigns a weak binding energy $-\eta<0$ to DNA tiles of species $\alpha$ and $\beta$ that are neighbors in both subsequent as well as preceding target structures, thereby ensuring reciprocity.  We define
\begin{equation}
    \Psi_{\alpha\beta}(d) = \begin{cases}
        -\eta, & \text{if there exists an $\ell$ for which} \\ & \text{ $\alpha\in S_\ell$ and $\beta\in S_{\ell+1}$ or} \\ & \text{ $\beta\in S_{\ell-1}$ are neighbors in}  \\ & \text{direction $d$;} \\ 
        0, & \text{otherwise},
    \end{cases}
\end{equation}
where the inter-target interaction energy $\eta$ is restricted to the range $[0,\epsilon]$ to ensure that these bonds remain weaker than the intra-target interactions.  One verifies that the interaction matrix $\Psi$ is reciprocal, satisfying $\Psi_{\alpha\beta}(d)=\Psi_{\beta\alpha}(-d)$.  

For the case of only two small target structures ($m=2$), each composed of only six distinct DNA tile species ($N_t=6$) from two non-overlapping libraries, a visual representation of the $12\times 12$ interaction matrices $U(d)$, ${\cal V}(d)$, and $\Psi(d)$ is provided in the Supplementary Information. The $d$-dependence is indicated by different colors, with red and green representing nonzero elements for $d=~\downarrow$ and $\uparrow$, respectively, and orange and yellow for $d =\ \rightarrow$ and $\leftarrow$, respectively.   

\subsection*{Swap rate}
\label{swaprate}
With the interaction matrices $U(d)$, ${\cal V}(d)$, and $\Psi(d)$ defined for a specific multi-component mixture of DNA tiles, tuned to stabilize  specific target structures and to facilitate transitions from one to the next structure, we  now introduce the MC swap rate for two tiles of species $\alpha$ and $\beta$ initially positioned at neighboring lattice sites $i$ and $j$, respectively.  This swap occurs with a probability
\begin{equation}
    p_{swap}(\alpha \leftrightarrow \beta) = \text{min}\left(1, \exp\left[{\frac{-\Delta \mathcal{H}+\Lambda}{k_BT}}\right] \right),
    \label{eq:swaprate}
\end{equation}
where $k_BT$ is the thermal energy unit at temperature $T$ and $\Delta \mathcal{H}$ is the swap-induced change of the (Hermitian, potential-based) binding Hamiltonian defined as
\begin{equation}
    \mathcal{H} = \sum_{\langle ij\rangle} s_i\cdot \Big(U(d_{ij}) + \Psi(d_{ij})\Big)\cdot s_j,
\end{equation}
where the summation is over all pairs of nearest neighbor sites $\langle ij\rangle$ of the cubic lattice. The non-reciprocity of the swap process is incorporated by the $\Lambda$-term in Eq.~(\ref{eq:swaprate}),  defined as
\begin{equation}\label{L}
\Lambda=s_i'\cdot\sum_{k} {\cal V}(d_{ik})\cdot s_k,
\end{equation}
where the sum over $k$ runs over the four nearest neighbor sites of site $i$, with directions $d_{ik}$. The sandwich of ${\cal V}(d_{ik})$ in Eq.~(\ref{L}) involves the post-swap state  $s_i'$ at site $i$. Specifically, if site $i$ becomes occupied by species $\beta$ after the swap, its  components are given by $s_{i,\mu}'=\delta_{\mu\beta}$. The states of the neighboring sites $s_k$ remain unchanged during the swap, so $s_k = s'_k$. If a $k$ site is occupied by a DNA tile of species $\gamma_k$, then $s_{k,\mu}=\delta_{\mu\gamma_k}$, whereas $s_{k,\mu}=0$ if it is unoccupied. 

In fact, $\Lambda$ as defined in Eq.~(\ref{L}) can  take  only five possible values, namely $\Lambda\in\{0,1,2,3,4\}\times\lambda$.

In the equilibrium limit $\lambda\to 0$, the non-reciprocal interactions vanishes, ${\cal V}(d)=0$, and the swap rate in Eq.~(\ref{eq:swaprate}) reduces to the standard Monte Carlo acceptance criterion based solely on the Boltzmann factor $\exp(-\Delta\mathcal{H}/k_BT)$ that guarantees detailed balance. \cite{Frenkel2023}  In contrast, for $\lambda>0$, detailed balance is no longer satisfied, and the dynamics becomes microscopically irreversible, in line with non-reciprocal  fuel consumption (or enzymatic activity) driving the system out of equilibrium.

\subsection{Monte Carlo simulations and system parameters}
\label{MC}
We perform canonical Virtual Move Monte Carlo (VMMC) simulations~\cite{Whitelam2007, Whitelam2009} on a three-dimensional discrete lattice of size 128$\times$128$\times$8 with periodic boundary conditions. In addition to standard (reciprocal) VMMC moves, we also incorporate Monte Carlo 
swap moves that induce non-reciprocal transitions between structures, with  acceptance rates defined by Eq~(\ref{eq:swaprate}).

In this work, we focus on two-dimensional  target structures composed of $18\times16=288$ DNA tiles, so that $N_t=288$ throughout this study. 
A typical configuration of a system with a fully assembled target structure is shown in Fig.~\ref{fig:phase-diagram}(a). In this example, we use $N_s=2N_t=576$ species, corresponding to a fraction of occupied sites  as low as $4.4\cdot 10^{-3}$. The interaction matrix $U(d)$ in Eq.~(\ref{U}) 
is designed to support two target structures $S_0$ and $S_1$, using non-overlapping libraries.  Here each species is assigned a unique color, chosen such that the fully self-assembled structures $S_0$ and $S_1$ resemble the paintings {\em Wheatfield with Cypresses} by  Vincent van Gogh and {\em The Milkmaid} by Johannes Vermeer, respectively. In the snapshot of Fig.~\ref{fig:phase-diagram}(a), $S_0$ is fully assembled, forming Van Gogh's painting, while the $N_t$ constituent tiles of $S_1$ remain dispersed throughout the simulation box, \textit{i.e.} they are not assembled into Vermeer's painting. Below we will see how to control which target structure forms and how to switch from one to the next.

We will also consider cases where multiple target structures are included with  (partially) shared particle libraries, \textit{i.e.} they use the same or largely overlapping sets of DNA tile species. To prevent the formation of chimeric structures in these scenarios, the target structures are designed to  create highly distinct local DNA tile environments. The algorithm used to enforce these distinct local DNA tile  environments in different target structures is based on the method of Evans \textit{et al.} \cite{Evans2024} and is described in detail below.

In this work, all simulations are performed at a bond strength  $\beta \varepsilon=6.8$, where $\beta^{-1}=k_BT$. At this value,  fully assembled target structures correspond to stable energy minima, while spontaneous nucleation of secondary structures from the fluid is suppressed, occurring only on timescales much longer than those accessible by our simulations. 
To enable transitions between target structures under these conditions, Osat {\em et al.} introduced non-reciprocal interactions that actively promote the nucleation of the next target structure within the current one, driven by a  consumable energy input $\lambda$ (as discussed above). 
To address the realistic case of Brownian motion, we introduce  additional stabilizing  bonds of strength $\eta$ between suitably selected tiles in both subsequent and preceding target structures. As we will show below,  these $\eta$-bonds, when chosen with appropriate strength,  prevent  fragmentation of intermediate structures during transitions driven under diffusive dynamics.

We perform simulations for a range of stabilizing interaction strengths $\beta \eta \in [0,5]$ and non-reciprocity values $\beta \lambda \in [0,5]$.  The target structures ($S_0$ and $S_1$), the number of species $N_s$ (and thus the number of tiles $N_t$) as well as the intra-bond interaction matrix $U$ are all identical to those used in Fig.~\ref{fig:phase-diagram}(a). The interaction matrix ${\cal V}$ is designed to facilitate  a non-reciprocal transition $S_0\rightarrow S_1$. 

Each simulation starts from an initial state closely resembling  the snapshot in Fig.~\ref{fig:phase-diagram}(a), with a fully assembled structure $S_0$ and the DNA tiles  that make up  $S_1$ freely dispersed in the surrounding fluid. 
We run VMMC simulations for a duration of $100\tau_0$, where the time unit is defined as $\tau_0=5\cdot 10^5$ MC sweeps. To determine the nucleation time $\tau_{nucl}$, we monitor the system at intervals of $\tau_0$ to check whether a nucleation event for the transition  to $S_1$ has occurred (defined as the moment where more than a third of the structure has transitioned). If  $S_1$ does not assemble within the simulation time, we assign $\tau_{nucl}=100\tau_0$.

In Fig.~\ref{fig:phase-diagram}(b) we show a heat map of $\tau_{nucl}/\tau_0$ on a logarithmic scale as a function of $\beta\eta$ and $\beta\lambda$. 

\section*{Discrete input,  non-reciprocity budget, and multifarious design}

Although the formulation of an efficient Potts-like lattice model and the identification of parameter values that enable the irreversible transition $S_0\rightarrow S_1$ between two Brownian target structures is a significant initial step, it remains only a first starting point toward the realization of a Brownian finite-state machine. Apart from the relatively straightforward extension to include libraries of DNA tiles that can self-assemble into  more than two target structures $S_\ell$, at least three additional nontrivial challenges must be resolved. These challenges involve
(i) 
designing transition sequences $S_\ell \to S_k$ of the finite-state machine in response to an external input sequence,
(ii) ensuring that input-dependent  transitions $S_\ell\rightarrow S_{k}$ proceed to full completion and from the desired $\ell$, while  (iii) suppressing unintended and premature transitions. Our strategy to tackle these challenges is a combination of introducing {\em discrete} Monte Carlo time intervals, defining a {\em time-dependent} non-reciprocity parameter $\lambda(t)$,  and employing {\em alternating libraries} of multifarious DNA tile species for consecutive target structures.

Our MC simulations always begin from the fully assembled target structure $S_0$ at time $t=0$. We define a fixed time interval $\Delta T = 5 \cdot 10^7 \text{~MC~sweeps} = 100 \tau_0$. This interval is sufficiently long  to ensure with high probability, under the present parameter settings, that a single transition $S_0\rightarrow S_1$ is completed after the trigger,  as shown above. 
We address challenge (i)  by allowing transitions $S_\ell \to S_k$ to be turned on or off depending on a binary input ``1'' or ``0'' . Physically, this trigger may correspond to the (de)activation of enzymes that facilitate these transitions. 
We distinguish transitions triggered by binary input ``1'' and those triggered by ``0''. 
Input signals are started at discrete MC times $t_n=n\Delta T$ for integer $n\geq 0$, 
when the initial magnitude of the non-reciprocal energy to stimulate a transition $S_\ell \to S_k$ is set to 
\begin{equation}
    \lambda_{\ell k}(t_n) = \begin{cases}
        \lambda_0, & \text{if the external input pulse at $t_n$} \\ & \text{enables the transition $S_\ell\to S_k$}; \\
        0, & \text{otherwise,}
    \end{cases}
    \label{lamt}
\end{equation}
where the nonzero default fuel energy is set to $\beta\lambda_0=2.9$. From the results in Section \ref{MC} and Fig.~\ref{fig:phase-diagram}(b), we expect that a structure $S_\ell$ completes a non-reciprocal transition to $S_k$ during the time interval $t\in[t_n,t_n+\Delta T]$, provided that  $\lambda_{\ell k}(t_n)=\lambda_0$. Conversely, setting $\lambda_{\ell k}(t_n)=0$ prevents the  transition during this interval. 
Note that input pulse ``0'' does not necessarily imply that all $\lambda_{\ell k}$'s are zero. 
For instance, in the case of finite-state automata for computing the modulo and for pattern recognition, the  input pulse ``0''  enables  a set of transitions, and input pulse ``1'' another set.  

In practice, however, when  $\lambda_{\ell k}(t_n)=\lambda_0$ is used to trigger the desired transition $S_\ell\rightarrow S_{k}$, it is essential to gradually reduce $\lambda_{\ell k}(t)$ during $t\in[t_n,t_n+\Delta T]$ to prevent additional undesired  transitions such as premature progression to the next structure that is accessible when $\lambda_{\ell k}(t)\neq 0$. A gradual reduction in non-reciprocity is physically plausible in systems composed of nucleic acids, where such interactions are mediated by fuel-consuming reactions powered for instance by ATP. In this case,  fuel availability is inherently limited by the slow diffusion of ATP  during a transition. To capture this constraint while preserving the simplicity of our Potts-like model, we introduce a (dimensionless) non-reciprocity ``budget'' $\mathcal{B}$, which represents the number of available fuel units $k_BT$ that can be expended within each time window $\Delta T$. As fuel is consumed, we assume that $\lambda_{\ell k}(t)$ decreases proportionally to the cumulative energy expended on these non-reciprocal processes, i.e. we write 
\begin{equation}\label{B}
    \lambda_{\ell k}(t) =  \left(1 - \frac{b(t)}{\mathcal{B}} \right) \lambda_{\ell k}(t_n) \text{ \,\,\,\,\,\,\,for } t\in[t_n,t_n+\Delta T],
\end{equation}
where $b(t)$ denotes the cumulative  energy spent (in units of $k_BT$) since the beginning of the time window, with $b(t_n)=0$. 
The decay of $\lambda_{\ell k}(t)$ should be slow enough (i.e. a sufficiently large budget ${\cal B}$) to allow completion of the desired transition to the next  target structure, yet fast enough (i.e. a sufficiently small  ${\cal B}$) to suppress any subsequent, unintended  transitions.

Even with a time-dependent reduction of  non-reciprocity during the time interval $\Delta T$, achieving a well-controlled transition becomes increasingly difficult in systems with more than two target structures. The challenge arises from the stochasticity of the desired transition $S_\ell\rightarrow S_{\ell +1}$, which in some cases occurs so quickly after  the trigger that sufficient budget remains to initiate the (undesired) next transition $S_{\ell+1}\rightarrow S_{\ell+2}$. To mitigate this, we 
implement a design rule in which target structures $S_{\ell}$ and $S_{\ell+2}$ are multifariously assembled from species belonging to (largely) overlapping libraries. This ensures  that the building blocks required for assembling $S_{\ell +2}$ are already incorporated in  $S_{\ell}$ and thus  depleted from the  solution as long as $S_{\ell}$ remains largely intact. By implementing this design principle of alternating (and overlapping) particle libraries along any path in the transition graph, we significantly reduce the likelihood of premature nucleation of subsequent target structures, specifically in the low-concentration regime that we consider here, where only a single copy of each  DNA tile species is present in the simulation box.

\section{Supplemental Material}
\begin{figure*}[t!]
    \centering
    \includegraphics[width=\linewidth]{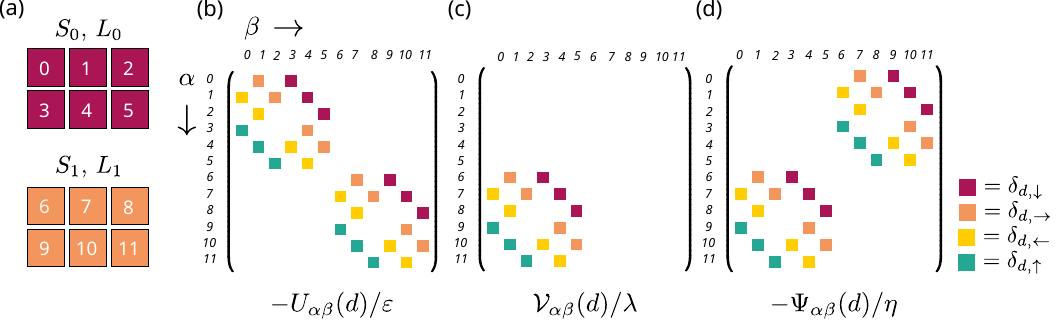}
    \caption{A schematic representation of the interaction matrices belonging to a system with two ($2\times 3$) target structures with a non-reciprocal transition. (a) The considered target structures $S_0$ and $S_1$, with libraries $L_0$ and $L_1$. They do not share a particle library so there are 12 particle species in total. The particle species label is shown inside the squares representing the individual particles.  (b) The $12\times12$ matrix $-U_{\alpha\beta}(d)/\varepsilon$ encoding the internal bonds of the target structures. (c) The matrix $\mathcal{V}_{\alpha\beta}(d)/\lambda$ encoding the non-reciprocal interactions. (d) The matrix $-\Psi_{\alpha\beta}(d)/\eta $ encoding the stabilizing bonds of intermediate structures. Only non-zero values of the matrices have been shown.}
    \label{fig:matrixsketch}
\end{figure*}
\begin{figure}[h!]
    \centering
    \includegraphics[width=\linewidth]{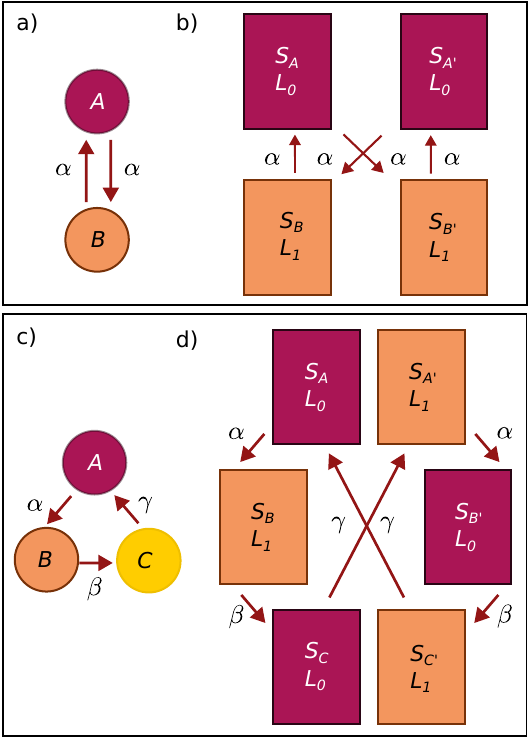}
    \caption{(a) Segment of a finite-state automaton that has two states $A$ and $B$, and two opposing transitions between them, both with input label $\alpha$. (b) Implementation of this segment in our system with multifarious target structures. Here we have four target structures $S_A$, $S_{A'}$, $S_B$, and $S_{B'}$. (c) Segment of a finite-state automaton with a cycle of 3 states, connected by transitions $\alpha$, $\beta$ and $\gamma$. (d) Implementation of this segment in our system with non-reciprocal transitions between multifarious target structures. Here there are 6 target structures, sharing 2 particle libraries}
    \label{fig:doublingrule}
\end{figure}
\subsection*{Interaction matrices}
An example of the interaction matrices, as defined in the main text, are illustrated in Fig.~\ref{fig:matrixsketch}. In this exemplary case, there are two small target structures $S_0$ and $S_1$, shown in Fig.~\ref{fig:matrixsketch}(a), composed of 6 unique particle species each, with a non-reciprocal transition $S_0 \rightarrow S_1$. In Fig.~\ref{fig:matrixsketch} the binding matrix $U_{\alpha\beta}(d)$ is shown, representing the internal bonds within the target structures. Hence, only the on-diagonal $6\times 6$ blocks contain non-zero entries. The squares in these matrices represent Kronecker delta functions of the direction of the interaction $d$ and the desired directions $\{\rightarrow, \downarrow, \leftarrow, \uparrow\}$, and encode the directionality of each designed bond. This matrix is reciprocal and obeys $U_{\alpha\beta}(d) = U_{\beta\alpha}(-d)$. The same holds for $\Psi$, where $\Psi_{\alpha\beta}(d) = \Psi_{\alpha\beta}(-d)$, which represents stabilizing bonds between particles that have neighboring locations, but belong to subsequent target structures. The interactions are therefore on the off-diagonal blocks. It is clear however, that the same reciprocity does not hold for $\mathcal{V}$, which only gives a (non-reciprocal) interaction between particles from target structure $S_1$ to those of $S_0$, but not vice versa. It is quite intuitive from this picture why it is not possible to design a non-reciprocal transition $S_0 \rightarrow S_1$ at the same time as a transition from $S_1 \rightarrow S_0$, as this would effectively transform the non-reciprocal interaction $\mathcal{V}$ into a reciprocal interaction, similar in shape to $\Psi$.

\subsection*{Multifarious design}
We make use of multifarious target structures in this work, meaning that most particles of one target structure are reused in other target structures. The algorithm with which we designed the target structures has been inspired by Evans \textit{et al.},\cite{Evans2024} and is designed to limit the number of chimeric structures that may form. 

We consider $N_{str}$ target structures that are intended to share a common particle library. Initially, each of the $N_{str}$ target structures possesses its own distinct set of particle species (288 in our case) and does not yet share any particles with the others. For each particle species, we identify which edges are inert (for example, those located at the boundaries or corners of the target structure), and therefore should not interact with particles belonging to other target structures. We then repeat the following procedure $10^4$ times. Two particle species, $i$ and $j$, are randomly selected from target structures $S_i$ and $S_j$, respectively, such that they do not yet appear in the same target structure and share the same set of inert edges. If species $i$ already appears in multiple target structures, one of them is randomly chosen as $S_i$. Species $j$ will replace species $i$ in structure $S_i$ if four criteria are satisfied.
First, the number of target structures containing species $i$ must be less than or equal to the number of structures containing species $j$. This condition ensures convergence of the algorithm.

Second, we must prevent any configuration in which an incorrect particle species could form two or more unintended bonds. If $j$ is inserted into $S_i$ in place of $i$, no other species should be able to form multiple incorrect bonds with neighboring sites. In other words, particles that interact with $j$, because they are adjacent in another target structure, must not be able to form a bond with any of the second neighbors of $i$ in $S_i$, as this could stabilize an incorrectly bonded particle.

Third, we extend this restriction one step further. If $j$ is inserted into $S_i$ and $j$ can bind to another species $k$, then there must be no species $l$ that interacts both with $k$ and with a third neighbor of $i$. Otherwise, this interaction could incorrectly stabilize the $k$–$l$ pair within $S_i$.

After completing this procedure, the $N_{str}$ target structures share a common particle library that is not much larger than the number of particle species necessary to form a single target structure. All particle species are then relabeled, and unused particle species are removed.

\subsection*{Network design}
As briefly mentioned in the main text, there are some key differences between finite-state automata and our system of multifarious structures with non-reciprocal interactions. In particular, our system is subject to two main constraints when  designing the network of target structures and transitions. First, we cannot implement opposing non-reciprocal transitions between two target structures $A$ and $B$ that share the same input label $\alpha$, as illustrated in Fig.~\ref{fig:doublingrule}(a). When such a configuration appears in the desired finite-state machine, it must be replaced by  a duplicated segment, as shown in Fig.~\ref{fig:doublingrule}(b). In this modified scheme, we eliminate opposing transitions between two target structures $S_A$ and $S_B$ by introducing duplicated states $S_{A'}$ and $S_{B'}$, redirecting one of the transitions to the duplicated branch. The duplicated states share the same particle libraries as their originals, i.e.  $S_A$ and $S_{A'}$ use the same library $L_0$, and $S_B$ and $S_{B'}$ share $L_1$.

Another constraint imposed by our design rules is that all possible transition paths within each time window must have alternating particle libraries. One important consequence of this requirement is that it is not possible to design certain odd-numbered cycles. A segment of a finite-state automaton that has such a cycle is illustrated in Fig.~\ref{fig:doublingrule}(c), where we have three states $A$, $B$ and $C$, that are connected by  transitions labeled $\alpha$, $\beta$ and $\gamma$. In our system, this segment is only possible with just three states if  $\alpha\neq\beta\neq\gamma$. If  two of the transitions share the same  input label, then the condition that in each time window all possible paths must have  alternating particle libraries cannot be met, as the cycle has an odd-numbered   length. In these cases, it is again possible to duplicate this segment by introducing $S_{A'}$, $S_{B'}$ and $S_{C'}$ and having one of the transitions of the cycle connected to the duplicated segment, as shown in Fig. \ref{fig:doublingrule}(d). We observe here that the duplicated target structures do not share the same particle library as their original counterpart due to the odd length of the cycle that is duplicated.

It should be noted that this duplication procedure does not generally require introducing additional particle species into the system, thanks to its multifariousness. Instead, the added  complexity arises from the design of the interactions.

\end{document}